# Maturity Model for IT Service Outsourcing in Higher Education Institutions

Victoriano Valencia García
Computer Management Technician
and Researcher at Alcalá University
Madrid, Spain

Dr. Eugenio J. Fernández Vicente
Professor at Computer Science Dept.
Alcalá University
Madrid, Spain

Dr. Luis Usero Aragonés
Professor at Computer Science Dept.
Alcalá University
Madrid, Spain

*Abstract*—The current success of organizations depends on the successful implementation of Information and Comunication Technologies (ICTs). Good governance and ICT management are essential for delivering value, managing technological risks, managing resources and performance measurement. In addition, outsourcing is a strategic option which complements IT services provided internally in organizations.

This paper proposes the design of a new holistic maturity model based on standards ISO/IEC 20000 and ISO/IEC 38500, the frameworks and best practices of ITIL and COBIT, with a specific focus on IT outsourcing.

This model is validated by practices in the field of higher education, using a questionnaire and a metrics table among other measurement tools. Models, standards and guidelines are proposed in the model for facilitating adaptation to universities and achieving excellence in the outsourcing of IT services. The applicability of the model allows an effective transition to a model of good governance and management of outsourced IT services which, aligned with the core business of universities (teaching, research and innovation), affect the effectiveness and efficiency of its management, optimizes its value and minimizes risks.

*Keywords— IT Governance; IT Management; Outsourcing; IT Services; Maturity Models*

## I. INTRODUCTION

One thing to change about ICT at university level is the deeply rooted approach which exists, or which used to exist, called infrastructure management. This kind of management has evolved into a governance and management model more in line with the times, which is a professional management of services offered to the university community [6]. It is for this reason that in recent years a set of methodologies, best practices and standards, such as ITIL, ISO 20000, ISO 38500 and COBIT, have been developed to facilitate ICT governance and management in a more effective and efficient way.

These methodologies, which are appropriate and necessary to move from infrastructure management to service management, see a lack of academic research. For that reason it is inadvisable to use these frameworks on their own, and it is advisable to consider other existing frameworks in order to extract the best from each for university level [6].

ICT or IT services have implications for business and innovation processes and may be a determinant in their evolution. The organization of these services, their status within the organization of the university, and their relationships with other management departments and new technologies is therefore vital. At present, the degree of involvement, the volume of services offered, and the participation or external alliances with partner companies through outsourcing are of special interest.

Currently, and in the years to come, organizations that achieve success are and will be those who recognize the benefits of information technology and make use of it to boost their core businesses in an effective strategic alignment, where delivery of value, technology, risk management, resource management, and performance measurement of resources are the pillars of success.

It is necessary to apply the above-mentioned practices through a framework and process to present the activities in a manageable and logical structure. Good practice should be more strongly focused on control and less on execution. They should help optimize IT investments and ensure optimal service delivery. IT best practices have become significant due to a number of factors, according to COBIT (www.itgi.org)[9]:

- Business managers and boards demanding a better return from IT investments, i.e., the demand that IT delivers what the business needs to enhance stakeholder value;

- Concern over the generally increasing level of IT expenditure;

- The need to meet regulatory requirements for IT controls in areas such as privacy and financial reporting, and in specific sectors such as finance, pharmaceutical and healthcare;

- The selection of service providers and the management of service outsourcing and acquisition;

- Increasingly complex IT-related risks, such as network security;

- IT governance initiatives that include the adoption of control frameworks and good practices to help monitor and improve critical IT activities to increase business value and reduce business risk;

- The need to optimise costs by following, where possible, standardised, rather than specially developed, approaches;

- The growing maturity and consequent acceptance of well-regarded frameworks, such as COBIT, IT Infrastructure Library (ITIL), ISO 27000 series on information security-related standards, ISO 9001:2000 Quality Management Systems—Requirements Capability Maturity Model ® Integration (CMMI), Projects in Controlled Environments 2 (PRINCE2) and A Guide to the Project Management Body of Knowledge (PMBOK); and

- The need for organizations to assess how they are performing against generally accepted standards and their peers (benchmarking)

It is clear that ICTs have become ubiquitous in almost all organizations, institutions and companies, regardless of the sector to which they belong. Hence, effective and efficient ICT management to facilitate optimal results is necessary essential.

Furthermore, in this environment of total ICT dependency in organizations using ICTs for the management, development and communication of intangible assets, such as information and knowledge [14], organizations become successful if these assets are reliable, accurate, safe and delivered to the right person at the right time and place [9].

In short, we propose that the proper administration of ICT will add value to the organization, regardless of its sector (whether social, economic or academic) and will assist it in achieving its objectives and minimizing risk [6].

Given the importance of proper management of ICT, the search for solutions to the alignment of ICT with the core business of organizations has accelerated in recent years. The use of suitable metrics or indicators for measurement and valuation, generate confidence in the management teams. This will ensure that investment in ICTs generates the corresponding business value with minimal risk [6].

The above solutions are models of good practice, metrics, standards and methodologies that enable organizations to properly manage ICTs. And public universities are not outside these organizations, though they are not ahead. In addition, interest in adopting models of governance and management of appropriate ICTs is not as high as it should be.

Two of the factors through which IT best practices have become important is, the selection of appropriate service providers and the management of outsourcing and procurement of IT services.

In addition, a maturity model is a method for judging whether the processes used, and the way they are used, are characteristic of a mature organization [4].

Models by phases or levels allow us to understand how IT management strategies based on computing evolve over time [10]. According to these models, organizations progress through a number of identifiable stages. Each stage or phase reflects a particular level of maturity in terms of IT use and management in the organization.

There are many maturity models in the literature, and they are applied to various fields, such as project management, data management, help desk, systems safety engineering. Most of them refer to either Nolan's original model [13] or the Capability Maturity Model of Software Engineering Institute (Carnegie Mellon Software Engineering Institute). The latter model describes the principles and practices underlying software processes and is intended to help software organizations evolve from ad-hoc, chaotic processes to mature, disciplined software processes.

Nolan was the first to design a descriptive stage theory for planning, organizing and controlling activities associated with managing the computational resources of organizations. His research was motivated by the theoretical need for the management and use of computers in organizations. From 1973 until today, technology and the way it is used has changed a lot, but Nolan's original idea is still valid, and it will remain so as long as the quality of services provided internally in organizations, or by external suppliers, are essential.

II. LITERATURE REVIEW ON MATURITY MODELS FOR IT OUTSOURCING AND COMPARISON CHART

Very few models or frameworks of IT outsourcing can be found in the literature, either from the point of view of the client or outsourcer. The few models or frameworks that exist are varied. After a thorough literature review, and taking into account the point of view of the customer, the following models have been found to be relevant:

- **[M1] Managing Complex IT Outsourcing – Partnerships (2002)** [2] focuses on managing complex relationships with IT vendors taking into account the following aspects: IT strategy; information management; flexible contracts; contract management and accounts; and human resource availability.

- **[M2] Information Technology Outsourcing (ITO) Governance: An Examination of the Outsourcing** Management **Maturity Model (2004)** [4] is an evolutionary model similar to CMM but it lacks metrics to measure the maturity level properly. The maturity model for IT outsourcing consists of 5 levels: level 1 (vendor management fundamentals); level 2 (defined service outcome); level 3 (measures); level 4 (trust); and level 5 (business value recognized).

- **[M3] A Unified Framework for Outsourcing Governance (2007)** [5]: a unified framework on the governance of outsourcing from the combined perspectives of the customer and the provider. The framework focuses on three areas: governance processes; organizational structure of governance; and performance measurement.

- **[M4] IT Outsourcing Maturity Model (2004)** [1]: this model identifies five maturity levels based on a literature review and interviews with the participants of outsourcing. It is a generalist, stage-maturity model where organizations rise gradually. The five stages or phases of the model are: insourcing; forming; storming; norming; and performing.

- **[M5] Outsourcing Management Framework Based on ITIL v3 Framework (2011)** [11]: a framework based on ITIL v3 that it is composed of four phases to be met consecutively. The four phases are: phase 1 (incident and problem management); phase 2 (financial management, risk management and service level management); phase 3 (supplier management; change management; and security management); and phase 4 (service evaluation).

- **[M6] Multisourcing Maturity Model (2011)** [8]: a multisourcing maturity model for large companies with federal IT organization. For that reason it is considered to be a very specific model. Model based on CMMI-ACQ, eSCM-CL and Gartner IT procurement, consisting of six levels: level 0 (multisourcing incomplete); level 1 (multisourcing prepared); level 2 (multisourcing engaged); level 3 (multisourcing established); level 4 (multisourcing managed); and level 5 (multisourcing optimized).

- **[M7] Maturity model for IT outsourcing relationships (2006)** [7]: a maturity model based on organizational theories and practices in the relationships established in IT outsourcing. The model consists of three phases: cost; resources; and alliances.

- **[M8] IT Governance Maturity and IT Outsourcing Degree: An Exploratory Study (2007)** [3]: the aim of this study is to shed light on whether the evaluation of IT governance maturity differs depending on whether clients outsource selectively or completely.

- **[M9] Global Multisourcing Strategy: Integrating Learning From Manufacturing Into IT Service Outsourcing (2011)** [12]: a theoretical framework with a specific focus on economic and operational outsourcing, it omits many other determinants. It proposes two dimensions –width and depth– which analyze the compensation of a multisourcing strategy in detail from the perspective of the customer.

The following table shows the maturity models and frameworks above, along with the key areas or determinants that they are based on. All key areas shown in the Table I are the bases of the maturity model designed for IT service outsourcing.

Taking into account all key areas shown in Table I, a holistic maturity model (henceforth MM) has been designed with a specific focus on IT outsourcing governance and IT service management.

The model establishes where organizations involved in the study are in relation to the following control criteria and information requirements according to Cobit: effectiveness; efficiency; confidentiality; integrity; availability; compliance; and reliability. Other criteria, from the perspective of managing critical IT resources; included: applications; information; infrastructure; and people.

TABLE I. EXISTING MATURITY MODELS AND FRAMEWORKS ON IT OUTSOURCING

| Key areas or determinants | Maturity models and frameworks about IT outsourcing | | | | | | | | |
|---|---|---|---|---|---|---|---|---|---|
| | *M1* | *M2* | *M3* | *M4* | *M5* | *M6* | *M7* | *M8* | *M9* |
| Formal Agreement | X | | | X | | X | X | | X |
| Service Measurement | | X | X | | X | X | | X | |
| Quality Management | | X | | | | | X | | |
| Monitoring and Adjustments | | X | X | | X | X | | X | |
| Alignment IT-Business | X | X | X | | | | | X | |
| IT Governance Structure | X | | X | | | | | X | |
| Service Level Agreement (SLA) | X | X | | X | X | | X | | |
| IT Service Registration | | | | | | | | | |
| Incident and Problem Management | | | | X | X | | | | |
| Changes | | | | | X | | | | |
| Testing and Deployment | | | | | | | | | |
| Control of External Providers | X | X | | | X | X | X | X | X |
| Business Risk | | X | X | | X | | | X | |
| Financial Management | | | | | X | X | X | X | X |
| Legislation | | | X | | X | | | | |
| Demand and Capacity Management | | | | | | | | | |
| Formal Agreement Management | X | | | | | | | | X |
| Knowledge Management | X | | X | | | | X | | |
| Guidelines on outsourcing an IT service (life cicle) | | | | | | | | | |

With regard to IT governance, standard ISO/IEC 38500:2008, published in 2008, aims to provide a framework of principles for directors of different organizations in order to manage, evaluate, and monitor the efficient, effective and acceptable use of information and communication technologies.

The direction, according to ISO / IEC 38500, must govern IT in three main areas:

- Management. Direct the preparation and implementation of strategic plans and policies, assigning responsibilities. Ensure smooth transition of projects to production, considering the impacts on the operation, the business and infrastructure. Foster a culture of good governance of IT in the organization.

- Evaluation. Examine and judge the current and future use of IT, including strategies, proposals and supply agreements (both internal and external).

- Monitoring. Monitor IT performance measuring systems in order to ensure that they fit as planned.

According to the results of the "IT Governance Study 2007" [15] [16], reasons compelling governments to create an IT structure in the university include: aligning IT objectives with strategic objectives; promoting institutional vision of IT; ensuring transparency in decision-making; cost reduction; increased efficiency; and regulation and compliance audits.

On service management, MM takes into account ISO/IEC 20000 and ITIL v3, but it is customized to integrate governance and management into a single model. The model moves towards an integration that facilitates the joint use of frameworks efficiently. Thus, the MM designed consists of five levels, with each level having a number of general and specific characteristics that define it. These are determined by the selection of general concepts that underpin the MM (see first column in Table I). The selection is always justified and countersigned by ISO 20000 and ISO 38500 standards and ITIL and COBIT best practice methodologies.

### III. MATURITY MODEL PROPOSED

In order to design the proposed maturity model, we studied in detail every reference on the provision of IT services that there is in the ISO 20000 and ISO 38500 standards and ITIL v3 and COBIT methodologies. In addition, we investigated the relevant literature and failed to find any maturity model that brings together the previous methodologies with a specific focus on IT outsourcing. As a result, a number of concepts and subconcepts were categorized to form the basis of the maturity model.

The MM follows a stage structure and has two major components: maturity level and concept. Each maturity level is determined by a number of concepts common to all levels.

Each concept is defined by a number of features that specify the key practices which, when performed, can help organizations meet the objectives of a particular maturity level. These characteristics become indicators, which, when measured, determine the maturity level.

The MM defines five maturity levels: initial or improvised; repeatable or intuitive; defined; managed and measurable; and optimized.

The model proposes that organizations under study should ascend from one level of maturity to the next without skipping any intermediate level. In practice, organizations can accomplish specific practices in upper levels. However, this does not mean they can skip levels, since optimum results are unlikely if practices in lower levels go unfulfilled.

The complete maturity model, with five levels and characteristics in each, is the subject of a paper selected to be published[1].

### IV. METRICS FOR MATURITY ASSESSMENT

We have designed an assessment tool along with the maturity model that allows independent validation and practical application of the model. Therefore, the maturity of an organization indicates how successfully all practices that characterize a certain maturity level have gone fulfilled. The questions used in the questionnaire consider the basis of the assessment instrument. They were extracted from each of the indicators defining each of the general concepts and key areas of the maturity model. These general concepts and defining characteristics have been extracted from the following standards and methodologies:

- Standard ISO/IEC 20000 and methodology of good practices ITIL v3. Both provide a systematic approach to the provision and management of quality IT services.

- Standard ISO/IEC 38500:2008 provides guiding principles for directors of different organizations to manage, evaluate, and monitor the use of information and communication technologies effectively and efficiently.

- Cobit business-oriented methodology provides good practice through a series of domains and processes, as well as metrics and maturity models in order to measure the achievement of the objectives pursued.

In addition, new indicators have been developed based on the proposed model in order to assess appropriate aspects not reflected either in previous methodologies and standards or in the existing literature (e.g. the inclusion of service performance in the SLA and the use of user-satisfaction surveys in IT-business alignment).

To evaluate the maturity model of an organization using the model and the measurement instruments proposed, it is necessary to obtain a series of data resulting from the responses to the questionnaire based on the indicators that define the general concepts of our maturity model.

Table II shows one of the nineteen key areas or concepts that are the basis of the MM. The first column of the table shows the level or levels corresponding to the indicator located in the second column.

The second column shows the survey questions and indicators for each of the questions or part of the questions. Finally, the third column shows the source where the indicator or item has been extracted as a feature of the general concept or key area of the model.

---

[1] Cf. Valencia, V., López, J., Holgado, J.C., & Usero, L., "Modelo de Madurez para la Externalización de Servicios de TI", Proceedings of the 8th International Academic Conference on Government and Management of IT Service, King Juan Carlos University, Madrid, [forthcoming]

TABLE II. METRICS TABLE AND QUESTIONNAIRE

| Level | Code – Indicator – Question of Questionnaire | Source |
|---|---|---|
| | **Concept: Formal Agreement: contract, agreement or similar (FA)** | ISO 20000, Cobit, ITIL v3 |
| 3 | **FA1** - Procedures and processes – Are there clear documented procedures to facilitate the control of outsourced IT services with clear processes for negotiating with external providers? | Cobit |
| | **FA2** - Elements of FA - Formal agreements (contracts, agreements or the like) of every outsourced IT service include: | ISO 20000, ITIL, Cobit |
| 3 | **FA2a** - Scope of work | |
| 2 | **FA2b** - Services / deliverables to be provided | |
| 3 | **FA2c** - Timeline | |
| 2 | **FA2d** - Service levels | |
| 2 | **FA2e** - Costs | |
| 3 | **FA2f** - Billing Agreements | |
| 2 | **FA2g** - Responsibilities of the Parties | |
| | **FA3** - Requirements of FAs - Formal agreements meet the following requirements: | Cobit |
| 3 | **FA3a** - Legal (compliance with current regulations) | |
| 3 | **FA3b** - Operational (proper delivery and management of services in operation) | |
| 3 | **FA3c** - Control (for the measurement and analysis of the services) | |
| 4 | **FA4** - Revision frequency of FAs - Formal agreements are reviewed periodically at predefined intervals | ISO 20000 |
| 3 | **FA5** - Penalties in FAs - There are penalties for breach of formal agreements, including termination of agreements | Self developed |
| 345 | **FA6** - Enforcement of penalties in FAs - Degree of enforcement of penalties for breach of agreements | Self developed |

Therefore, the maturity level of every higher education institution studied is measured by evaluating its development in each key area or concept, which is indicated by responses to items or indicators in Table II. In order to qualify for a specific maturity level, the university surveyed must carry out all key practices of that level successfully.

V. OBJECTIVES OF THE MATURITY MODEL

The main purpose of the model is to fulfill as many requirements of an ideal maturity model for IT outsourcing in the governance and management of outsourced IT services as possible. With the identification and definition of some key concepts and an assessment tool, the model allows a systematic and structured assessment of organizations. Although the assessment instrument has a lot of qualitative responses, it also has quantitative responses, such as the degree of compliance with certain characteristics that define the maturity model (e.g. the degree of influence of the KPIs and KGIs in the penalties for breach of agreements).

The identification of key areas and concepts specifying its characteristics to constitute the underlying structure of the MM, complements the necessity to refer to governance and management concepts tested and backed by standards and methodologies.

Moreover, the model advocates continuous learning and improvements in governance in IT outsourcing and good management of outsourced IT services, even when organizations have reached the maximum level (5).

VI. CONCLUSIONS AND FUTURE RESEARCH

Both ISO 20000 and ISO 38500 standards, and ITIL v3 and COBIT methodologies of best practice in IT management and governance, are a good basis for the study and analysis of governance and management of the outsourced IT services in organizations. That is why they allow the design of a new maturity model that facilitates the achievement of an effective transition to a model of good governance and management of outsourced IT services that, aligned with the core business in organizations, impacts on the effectiveness and efficiency of its management, optimizes its value and minimizes risks.

A questionnaire (survey form) forms the basis of the quantitative study of the maturity model. The questionnaire is based on the attributes or indicators that define the different levels of the model. It contains standard and suitable questions, according to the nature of the research.

Questionnaire responses allow the obtaining or calculation of the level of maturity by applying the scale defined in the model. In addition, questionnaire responses, after being properly analysed, shed light on the current situation of the different organizations studied in governance and management of outsourced IT services.

This research will also provide specific case studies that will be carried out at some universities and will put the model into practice in order to draw conclusions. The questions used in the questionnaire, currently sent to some universities to be completed, bring the design of a proposed improvement plan (see Table III) to allow a sequential growth by stages. The growth occurs as a hierarchical progression that should not be reversed, for the aforementioned reasons, and involve a broad range of organizational activities in governance and management of IT outsourcing.

Table III shows one of the five levels (there are five tables, one for each level) of the MM with the key areas or concepts to be improved in order to allow a sequential growth by stages. The first column of the table shows the concepts. The second column of the table shows the objectives to achieve corresponding to the concept in the first column. Finally, the third column shows the actions to accomplish in order to achieve the objectives set in the second column.

Therefore, in the case studies, we will apply the established scales, which will rate the university surveyed and the object of study, at a level of maturity within the MM. Depending on the level of maturity in which the university is rated, improvement actions, according to the improvement plan, will be proposed to achieve a target level.

TABLE III. IMPROVEMENT PLAN. LEVEL 1

| Level 1 - Initial or improvised |  |  |
|---|---|---|
| Concept | Improvement Objectives | Improvement Actions |
| Formal Agreement: Contract, agreement or similar (FA) | - Basic formal agreement (contract, agreement or similar) | - IT Management must understand the necessity to sign a basic contract, agreement or similar of IT services meant to be outsourced |
| Service Measurement | - If there is measurement (quality, performance, risks) of the IT services provided externally, it is informal and reactive | - The measurement of quality, performance and risks of outsourced IT services is essential to meet their expectations and business needs. In addition, measurement help early detection of potential problems. Therefore, it is advisable to do this measurement, even if it is informal and reactive. |
| Alignment IT-Business | - The requirements of the outsourced IT services are not defined, implemented and aligned with business objectives |  |
| IT Governance Structure | - There is not an organizational structure of IT Government where the CIO or equivalent is the backbone |  |
| Service Level Agreement (SLA) | - Basic SLA, if any, and without the following: responsibilities of the parties, penalties for breach of agreement, recovery time, levels of quality, security requirements and performance requirements of the service | - SLA is the reference document where is stated how the service signed between the service provider and the customer is provided. Therefore, it would be advisable to have a basic SLA with the essential aspects of the outsourced IT service, such as the service description and availability |
| IT Service Registration | - Basic service catalog without the following: terms of provision of services, SLAs, costs and responsibilities of the parties | - A simple service catalogue with basic information should be created. It would lack the following: terms of provision of services, SLAs, costs and responsibilities of the parties |
| Incident and Problem Management | - There is no optimized tool to manage incidents |  |
| Testing and Deployment | - Success depends on IT staff experience, and improvisation rules | - Improvisation should stay out of the IT operational deployment and testing, but since there is improvisation, IT staff (internal and external) and end users should be well trained because the success depends on them |
| Legislation | - Loopholes in data protection, data processing, location where data processing takes place, clauses for the transfer of data and standard contractual clauses for the transfer of personal data to third countries |  |

The measurement process to ascend in the MM is as follows (see Fig. 1):

*1) Perform an initial measurement after completing the questionnaire;*

*2) Set goals (benchmark);*

*3) Identify the gaps between the current measurement and the objectives set;*

*4) Recommend actions and policies to be implemented within the improvement plan to ascend in the MM; and*

*5) Once corrective actions have been implemented, perform a new measurement.*

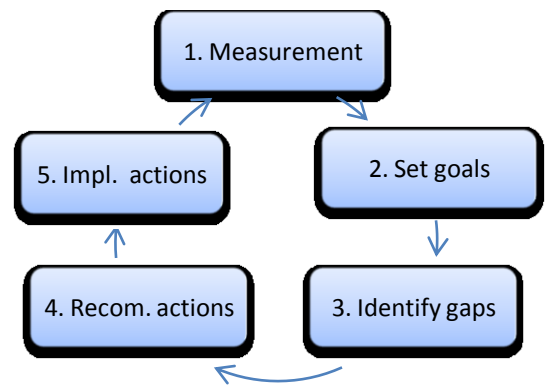

Fig. 1. Measurement process

In this context, the model is going to be validated in the field of higher education. Furthermore, models, standards and guidelines are recommended in order to enable and facilitate adaptation to universities so that they can move up the maturity model. Thus, the model, based on standards and best practices, is designed to achieve excellence in the management of IT outsourcing. The applicability of the study, by the case studies mentioned before, allows universities to meet the goal of effective transition to a model of good governance and good management of outsourced IT services. Aligned with the core business of universities (education, research and innovation) this will impact on the effectiveness and efficiency of their management, optimize value and minimize risks.

On the basis of this research, future studies will provide some conclusions and reflections on the future of IT outsourcing and other IT service delivery formulas in 21st century digital universities, in order to allow them to meet successfully the requirements of the European Education Higher Area (EEHA) and the complex digital era of the internet.